\documentclass[seceq]{ptptex}

\usepackage{graphicx}
\usepackage{wrapft}

\notypesetlogo                       

\title{
Dynamical properties of temperature chaos and memory effect%
}

\author{
Shu \textsc{TANAKA}$^{1,}$\footnote{E-mail: shu-t@spin.phys.s.u-tokyo.ac.jp}, and Seiji \textsc{MIYASHITA}$^{1,2}$%
}

\inst{
$^1$Department of Physics, School of Science, The University of Tokyo,\\
Tokyo 113-0033, Japan \\
$^2$Department of Applied Physics, School of Engineering, The University
of Tokyo, \\Tokyo 113-8656, Japan
}

\abst{

In the frustrated interaction systems, the nature of ordered
configuration can be intrinsically temperature dependent. There, the idea
of effective coupling of decorated and frustrated bond plays an
important role. The idea of effective coupling is verified in the study
of equilibrium properties. We will study how the effective coupling is
realized in the dynamical processes, e.g. the relaxation time
after a change of temperature. We also study the memory phenomena which
have been found in the spin glass.

}

\begin{document}

\maketitle

\section{Introduction}

Researches on the spin glass have been made in both-sides side of a
theory and an experiment for the past dozens of years. An interesting
dynamical characteristic called aging phenomena have
appeared\cite{rf:1}\tocite{rf:11}. There are following features in the
aging phenomena. The first is \textit{memory effect}. It is observed by
the experiments of temperature cycling\cite{rf:1}\tocite{rf:7}. In these
experiments, temperature change is carried out at a fixed speed, and at
a certain temperature $T_1$ the temperature is fixed for a time
$t_{\mathrm{w}}$(waiting time). During which the dynamical
susceptibility reduces to a certain value. After that, the temperature
is changed again up to $T_2$($T_1 > T_2$), and then it is swept back to
$T_1$ at the same speed. The value of susceptibility observed at $T_1$
is reproduced, this phenomenon is called memory effect. The second is
\textit{rejuvenation}. It is found in the same protocol where the memory
effect is observed. When the temperature is reduced below $T_1$, the
susceptibility increases again. This means that the system is more
disordered, although the temperature is decreased, which is called
\textit{rejuvenation}. This phenomenon indicates that ordering pattern
is temperature dependent, which is called \textit{temperature chaos}.

\par 

Although interesting phenomena have been observed in experiments, there
are only few explanations for these phenomena from microscopic
viewpoint. Yoshino et al.\cite{rf:15} and Miyashita et al.\cite{rf:12}
studied a mechanism of the memory effect. The latter introduced a
microscopic mechanism of the temperature dependence and an idea of
``memory spot'', by which they demonstrated the memory effect. But in
their study, the change of the effective temperature is assumed by
hand. Thus, in this paper, we will study a microscopic bond
configuration which realizes the change of the effective coupling
automatically, and try to realize the memory effect. We study a model
with a certain combination of bond which is called ``decoration
bond''\cite{rf:13}.

\section{Model and Numerical experiment protocol}

We introduce a decoration bond in order to realize temperature
dependence of the effective coupling, and study microscopic mechanism of
the dynamics of the memory effect. As shown in Fig. 1(a), we prepare a
system which is a $10 \times 10$ square lattice. Each bond of the
lattice, which we call ``system bond'', consists of a set of bonds
depicted in Fig. 1(b). The effective coupling of the bond
$K_{\mathrm{eff}}$ is defined by $e^{K_{\mathrm{eff}}\sigma _1 \sigma
_2} \propto \mathrm{Tr}_{\{ s \}} e^{-\beta \mathcal{H}}$. Concretely,
for Fig. 1(b), we have

\begin{equation}
 K_{\mathrm{eff}} = m\cdot \left(\frac{1}{2} \log \left[ \frac{\cosh \beta\left( J_1 + J_2 \right)}{\cosh \beta \left( J_1 - J_2 \right)}\right] - \beta J_3\right) \qquad \left( \beta = \frac{1}{k_{\mathrm{B}}T} \right),
\end{equation}
where $m$ is the number of sets of the decoration bond in a system bond.
\par

Because $K_{\mathrm{eff}}$ of this single decoration bond does not
 exceed, the transition value $K_{\mathrm{c}}$ of the two
 dimensional Ising model. For this reason, we arrange ten sets of the
 decoration bond(see Fig. 1(c)). The total effective coupling exceeds
 $K_{\mathrm{c}}$. So that, it is expected that ferromagnetic order
 appears at $T_1$ and antiferromagnetic order at $T_2$. We consider that
 an effective interaction contributes to the temperature selectivity
 nature of an aging phenomenon. Temperature dependence of
 $K_{\mathrm{eff}}$ is nonmonotonic. In other words, it shows a
 reentrant phenomenon\cite{rf:14}. Using $J_1 = 0.6$, $J_2 = -0.7$, $J_3
 = 0.28$ (positive constant denotes a ferromagnetic coupling), the
 effective coupling exceeds $K_{\mathrm{c}}$ as depicted in
 Fig. 1(d). For this reason, at $T_1$ the system is ordered
 ferromagnetically, and at $T_2$ the system is ordered
 antiferromagnetically. 
\par

However, in fact, at low temperature
 side($T_2$), the order expected from $K_{\mathrm{eff}}$ does not appear
 in a short time. A kind of freezing appears if the
 values of decoration bonds are not chosen appropriately. About this
 problem, we will discuss in the next section. Next, we introduce a
 configuration for a memory spot, which is drawn by the thick line in
 Fig. 1(a). It serves as a local strong interaction to keep memory.

\begin{figure}[h]
 \centerline{\includegraphics[height=5cm,clip]{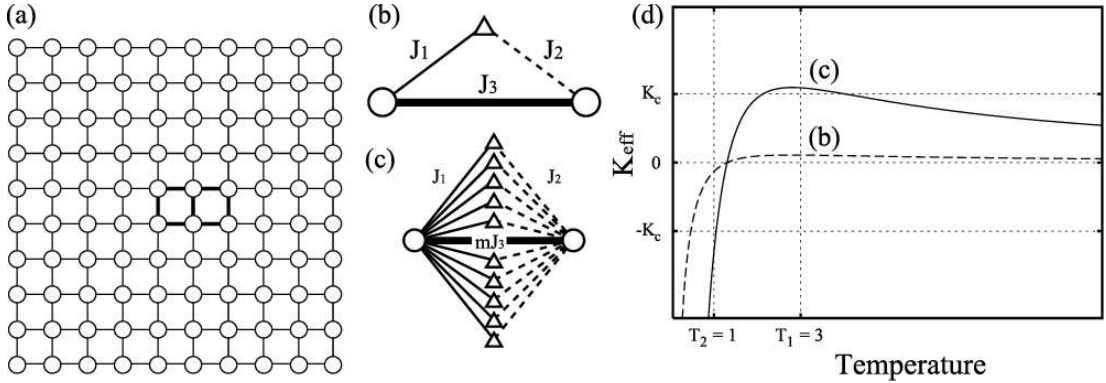}}
 \caption{(a) A $10\times 10$ lattice of two dimensional square lattice. (b) the decoration bond system. (c) a system bond(10 sets
 of the decoration bond (b)), (d) the effective coupling of the bond (c).}
 \label{}
\end{figure}

\section{Dynamical aspects and Results}
\subsection{Mechanism of slowing down}

\begin{wrapfigure}{r}{\halftext}
 \centerline{\includegraphics[height=3cm,clip]{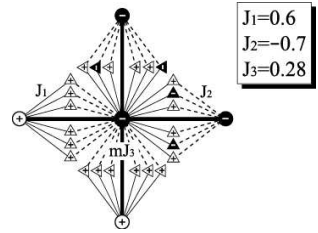}}
 \caption{Microscopic configuration of decoration spins.}
 \label{}
\end{wrapfigure} 
It turned out a sequence of ferromagnetic order $\to$ antiferromagnetic
order $\to$ ferromagnetic order is brought thanks to the effective
coupling which is induced by the decoration bonds. However, as pointed
out in the previous section, the expected order does not in a short
time. Fig. 2 shows a 0 of decoration spins which does not relaxed easily,
although the center spin can flip free by in the lattice of a single
bond. We defined an effective time $t_{\mathrm{eff}}$ by calculating a
probability of a flip of this ``free system spin''. In Fig. 2, between
the center spin ($-$) and the $+$ spins, the internal field on the
decoration spin is $J_1-J_2=J_1+|J_2|=1.3$, and the decoration spins are
ordered in + direction with a high probability
$p=\exp(1.3\beta)/(1+\exp(1.3\beta))$. The expectation value of the
internal field from the $+$ spins $H^+$ is $\langle H^+\rangle = m\times
(1-2p)(J_1+|J_2|)$. On the other hand, between the center spin and the
$-$ spins, the internal field on the decoration spin is
$-J_1-J_2=-J_1+|J_2|=0.1$, and the decoration spins are ordered in $+$
direction with a low probability $p'=\exp(0.1\beta)/(1+\exp(0.1\beta))$.
The expectation value of the internal field from the $+$ spins $H^-$
is $\langle H^-\rangle = -m\times (1-2p)(-J_1+|J_2|)$. Thus the internal
field on the center spin tends to be negative and the center spin is
stabilized to be the present value.  If the center spin is initially $+$,
the state is stabilized in the same scenario. The center spin ($-$) can
flip when $H^++H^- \le 0$, the probability of which is very small. We
defined an effective time $t_{\mathrm{eff}}$ by considering flipping
probability of this free system spin. In the present situation,
$t_{\mathrm{eff}}$ is estimated to be about 357 at $T_2$, which caused
slowdown of the relaxation by about 357 times. If we use the same
strength for $J_1$ and $J_2$, the ratio of $p$ and $p'$ becomes large
and $t_{\mathrm{eff}}\simeq 613$.  \par It should be noted that, if we
use a longer decoration bond, the domain wall presents near the center
of bond and the system spins are screened by favorable neighborhood
spins and strongly stabilized, which causes extremely long relaxation.

\subsection{Memory spot to realize the memory effect}

In this subsection, we will demonstrate this memory of a ordering
pattern. We consider a large system consisting of $5 \times 5$ of the
unit clusters. Fig. 1(a) coupled by a weak decoration bond($J' =
0.2J$). After a new order develops at $T_2$, the order at $T_1$ is
erased, and when the temperature comes back to $T_1$, the original order
is usually not reproduced. However, if the memory spot is introduced,
the order is reproduced. In Fig. 3, ferromagnetic order formed at
20000MCS. After that, temperature is changed to $T_2$. There, the
ferromagnetic order is destroyed, and antiferromagnetic order begins to
form. However, as we explains above, the development of
antiferromagnetic order is very slow. At 40000MCS, the antiferromagnetic
order completed at each cluster. And after the temperature is increased
to $T_1$. There, the antiferromagnetic order is destroyed and
ferromagnetic order appears in the same pattern as the previous
one. This process demonstrates the memory effect. In each cluster,
the ferromagnetic order spreads from each memory spot. It will be an
interesting future problem to find some design to realize two or more
memories.

\begin{figure}[h]
 \centerline{\includegraphics[height=6.5cm,clip]{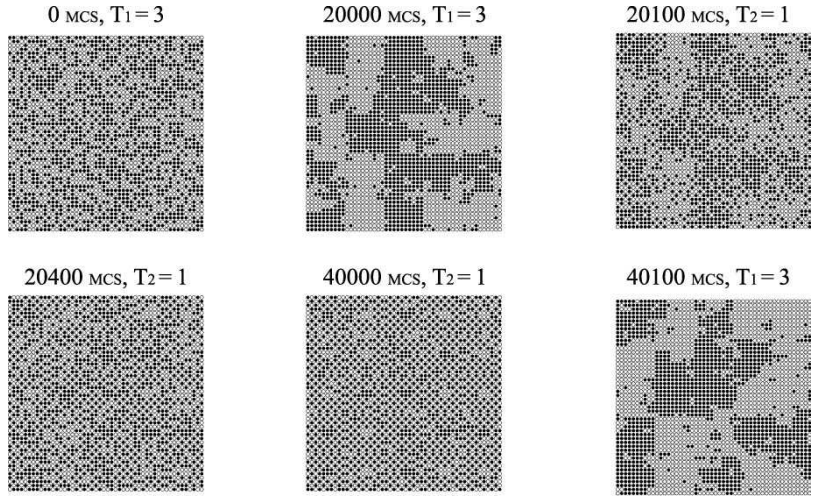}}
 \caption{From 0 to 20000 MCS, the temperature is $T_1 = 3$ at which
 $K_{\mathrm{eff}} > K_{\mathrm{c}}$, as a result, the ferromagnetic
 order appears. Next, From
 20000 to 40000 MCS, the temperature is $T_2 = 1$ at which
 $K_{\mathrm{eff}} < -K_{\mathrm{c}}$. As a result, the
 antiferromagnetic order appears. At last, the configuration returns to
 that at 20000 MCS due to the memory spot.}
 \label{}
\end{figure}

\section*{Acknowledgements}
The present work is partially supported by NAREGI Nanoscience Project,
Ministry of Education, Culture, Sports, Science and Technology, Japan.

%

\end{document}